\documentclass[prb,aps,twocolumn]{revtex4}

\usepackage{graphicx}        % standard LaTeX graphics tool
\usepackage{graphicx,psfrag,color}
\usepackage{amsmath,amssymb}

\usepackage[latin1]{inputenc}
\usepackage{url}

\begin{document}

\title{Signal transport in and conductance of correlated nanostructures}
%% for an abbreviated version of
\author{Peter Schmitteckert}
\affiliation{DFG Center for Functional Nanostructures, Karlsruhe Institute of Technology, 76128 Karlsruhe, Germany}
\affiliation{Institute of Nanotechnology, Karlsruhe Institute of Technology, 76344 Eggenstein-Leopoldshafen, Germany}
\begin{abstract}
Here we report on our project concerning
the application of time dependent DMRG to strongly
correlated systems. We show that a previously reported
simulation of the spin charge separation in a one-dimensional
Hubbard system exceeds a relative error of 100\% in the spin sector.
In the second part we discuss the application of the Kubo formula to
obtain linear conductance for the interacting resonant level model.
\end{abstract}

\maketitle

Transport properties of strongly interacting quantum systems
are a major challenge in  todays condensed matter theory.
While much is known for transport properties of non-interacting electrons,
based on the Landauer Büttiker formalism, the non equilibrium properties
of interacting fermions are an open problem.
Due to the vast improvements in experimental techniques there is an 
increasing theoretical interest in one-dimensional quantum systems.
Since in low dimension the screening of electrons is reduced the
effective interaction gets increased and can drive the electron systems
into new phases beyond the standard description of a Fermi liquid, 
e.g.\ into a Luttinger liquid. 

Formally the conductance of a quantum device attached to leads is given
by the Meir Wingreen formula. Besides the special case of proportional coupling,
the Meir Wingreen can only be treated within perturbative approaches.

The density matrix renormalization group method \cite{White92,White93,DMRGReviews}
is a well established method to treat onedimensional interacting quantum systems.
In this project we apply the real time evolution within the
density matrix renormalization group method (RT-DMRG)
to simulate the signal transport in onedimensional, interacting quantum systems,
and the conductance of interacting nanostructures attached to onedimensional,
non-interacting leads. 
In addition we calculate
the conductance from the current-current and current-density correlations functions
as a comparison to the real time evolution scheme and as a tool as itself, as it allows
for a higher energy resolution as compared to the real time approach.

In this project we developed a DMRG code applying Posix threads to parallelize the code
which is described in detail in \cite{Schneider_Schmitteckert:2006}.
While the DMRG is an approximative scheme, it has a systematic parameter, namely the number
of states kept per block, to increase the accuracy of the calculation.
In section \ref{sec:SC} we show that this code allows us to perform systematic
studies of the accuracy of transport problems. The major problem that arouse during our
previous work \cite{SS06,BSW06,S04,Schneider_Schmitteckert:2006} lies in the large
ressources needed to perform the actual simulation. In section \ref{sec:Kubo} we show
that we have now reformulated the Kubo approach which allows us to obtain a much higher energy
resolution and that we could get rid of numerical instabilities.

%%
%%
%%%%%%%%%%%%%%%%%%%%%%%%%%%%%%%%%%%%%%%%%%%%%%%%%%%%%%%%%
\section{Spin charge separation}
\label{sec:SC}
%%%%%%%%%%%%%%%%%%%%%%%%%%%%%%%%%%%%%%%%%%%%%%%%%%%%%%%%%
%%
%%
The spin charge separation of a single electron excitation is a prominent
example of interaction effects in onedimensional electron systems.
The first numerical observation was performed with an exact diagonalization approach
by Karen Hallberg et al.\ \cite{Jagla_Hallberg_Balseiro:1993} for a 16 site system.
Kollath et al.\  \cite{Kollath_Schollwoeck_Zwerger:2005} reported a simulation 
on a 72 site system with hard wall boundary conditions and 56 electrons.
In \cite{Schneider_Schmitteckert:2006} we showed that
with our code it is possible to study spin charge separation within the frame work of RT-DMRG
for a 2/3 filled  33 site Hubbard chain with periodic boundary conditions (PBC). 
The advantage of PBC lies is the absence of Friedel oscillations from the boundary.
It turned out that for accurate results we should at least use of the order of 2000 states per block, 
which is considerably more than applied in \cite{Kollath_Schollwoeck_Zwerger:2005}. 

Here we compare the results of Kollath et al.\ \cite{Kollath_Schollwoeck_Zwerger:2005} (KSZ)
who employed an adaptive RT-DMRG scheme combined with a Trotter decomposition
\cite{White_Feiguin:2004,Daley_Kollath_Schollwoeck_Vidal:2004}
 with results
obtained from our code \cite{Schneider_Schmitteckert:2006} where we combine the adaptive scheme
with a Krylov based matrix exponential \cite{S04}.
The system is a 72 site Hubbard model with an on site interaction of $U=4.0$.
The perturbation was created by applying a Gaussian perturbation to the potential of the
up-electrons in the same way as described in \cite{S04}.

\begin{figure}
 \centering
 \includegraphics[width=\columnwidth]{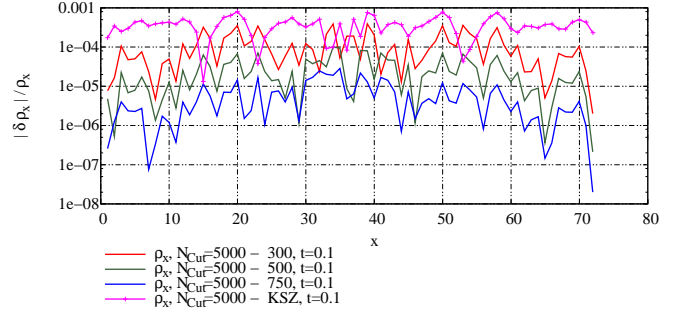}
 \caption{Comparison of the electron density $n(x)$ of KSZ and our calculations keeping 300, 500, and 750 states per block,
	$M=72$ sites, $N_\uparrow = N_\downarrow=28$
 with a reference calculation keeping 5000 states per block at time step t=0.1.}
 \label{fig:SC1}       % Give a unique label
\end{figure}

\begin{figure}
 \centering
 \includegraphics[width=\columnwidth]{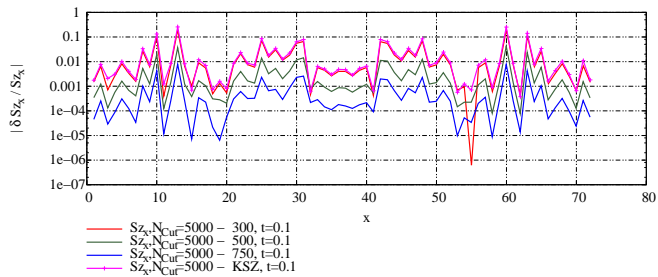}
 \caption{Comparison of $S^z(x)$ of KSZ and our calculations keeping 300, 500, and 750 states per block
 with a reference calculation keeping 5000 states per block at time step t=0.1.}
 \label{fig:SC2}       
\end{figure}

In Fig. \ref{fig:SC1} (\ref{fig:SC2}) we plot the relative accuracy of the electron density $n(x)$
(and its spin component $S^z(x) = (n_{\uparrow}(x) - n_{\downarrow}(x))/2)$ at time step $t=0.1$.
It shows that KSZ and a 300 state calculation already have a relative accuracy which exceeds $10^{-3}$
for the electron density, while for the spin component one has to go up to $750$ to achieve an
accuracy of the order of $10^{-3}$. The relative accuracy for the spin component is much harder as it
can get close to zero.

\begin{figure}
 \centering
 \includegraphics[width=\columnwidth]{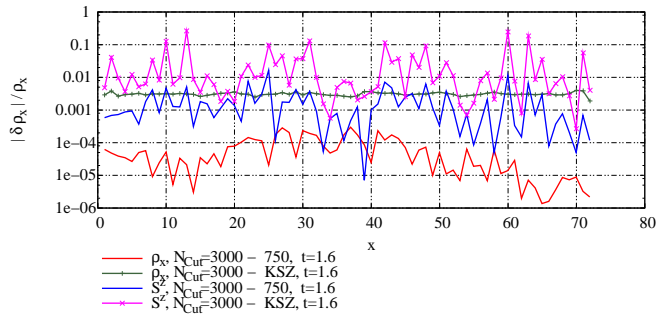}
 \caption{Comparison of the density $n(x)$ and $S^z(x)$ of KSZ and our calculations keeping 750 states per block
 with a reference calculation keeping 3000 states per block at time step t=1.6 after an initial run keeping 5000 states
up to $t=0.1$.}
 \label{fig:SC3}       
\end{figure}

After performing an initial calculation keeping 5000 states per block up to time $t=0.1$,
we continued the time evolution with 3000 states per block up to $t=12$.
In Fig. \ref{fig:SC3}  we compare our results with KSZ and a calculation keeping 750 states per block.
It shows that keeping 750 states per block we can still obtain an accuracy below 1\% for the density and
the spin component, while KSZ achieve and accuracy of 1\% only for the density, while the spin component 
goes above an error of 10\%.

Finally we compare the results of KSZ with our reference calculation keeping 3000 states per block
at time time step $t=11.6$. While KSZ are able to achieve an accuracy of 1\% for the absolute numbers,
the spin component shows a relative deviation larger than a few hundred percent.
While we have to be careful whether our results can be trusted at $t=11.6$
to serve as an accuracy benchmark, the calculation should be much more accurate than the one performed by KSZ.
In summary we have shown that one has to be very careful when employing the
real time extensions to the DMRG. However, DMRG allows for a systematic check of the results which is a very 
important property in a field where no other benchmarks are available.

\begin{figure}
 \centering
 \includegraphics[width=\columnwidth,clip]{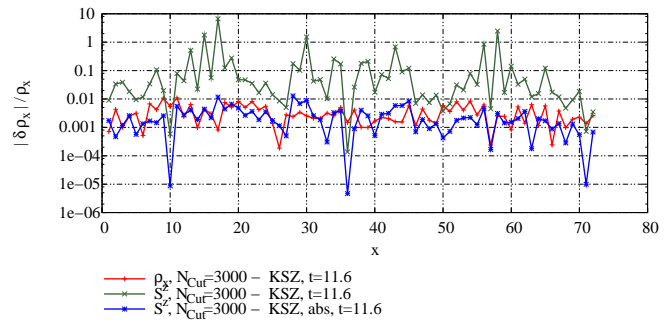}
 \caption{Comparison of the density $n(x)$ and $S^z(x)$ of KSZ 
 with a reference calculation keeping 3000 states per block at time step t=11.6 after an initial run keeping 5000 states
up to $t=0.1$. For the $S^z(x)$ component we plot the relative and the absolute difference.} 
 \label{fig:SC4}       
\end{figure}

%%
%%
%%%%%%%%%%%%%%%%%%%%%%%%%%%%%%%%
\section{Linear response with momentum leads}
\label{sec:Kubo}
%%%%%%%%%%%%%%%%%%%%%%%%%%%%%%%%
%%
Linear response calculations within DMRG \cite{BSW06} provide a method to
calculate the conductance of a nanostructure attached to leads.
As it is based on the the exact Kubo formula for the
linear conductance  $g\equiv
\frac{e^2}{h}\big<\tilde J\big>/V_{SD}$
it is valid for arbitrary interaction.
In the DC limit the conductance can be expressed in
terms of two different correlators,
\begin{eqnarray}
  g_{J_jN} &=& -\frac{e^2}{h}\big<\psi_0\big|\hat{J}_{n_j}
  \frac{4\pi i\eta}{(\hat H_0-E_0)^2+\eta^2}\hat N\big|\psi_0\big>,\label{g_JV definition}\\
  g_{JJ} &=& \frac{e^2}{h}\big<\psi_0\big|\hat{J}_{n_1}\frac{8\pi \eta(\hat H_0-E_0)}{\big[(\hat H_0-E_0)^2+\eta^2\big]^2}\hat{
  J}_{n_2}\big|\psi_0\big>,\label{g_JJ definition}
\end{eqnarray}
where the positions $n_j$ are in principle arbitrary. However, the positions
$n_1$ and $n_2$ should be placed close to the nanostructure to minimize finite size effects.
Bohr, Wölfle and Schmitteckert \cite{BSW06} had to introduce exponentially reduced hopping terms
close to the boundary of the leads which had been described in real space  to minimize finite size effects, 
which in return leads to ill-conditioned linear systems.
In order to solve these equations, they had to employ scaling sweeps to switch on the damping in the leads gradually.
While the method proofed to be a valuable tool it turned out that it is getting too expensive to study more
interesting systems.

Recently we have developed a new scheme \cite{Bohr_Schmitteckert:2007} based on leads described 
in momentum space to overcome
the difficulties we encountered in \cite{BSW06}, for details see also \cite{Schneider_Schmitteckert:2006}. 
While it is generally accepted that DMRG
does not work well in a momentum space description due to the large amount of couplings
intersecting the artificial cut of the system into two parts within DMRG, our transport
calculation are performed with non interacting leads. Therefore the number of links intersecting
the DMRG splitting of the system is vastly reduced.

\begin{figure}
 \centering
 \includegraphics[width=\columnwidth]{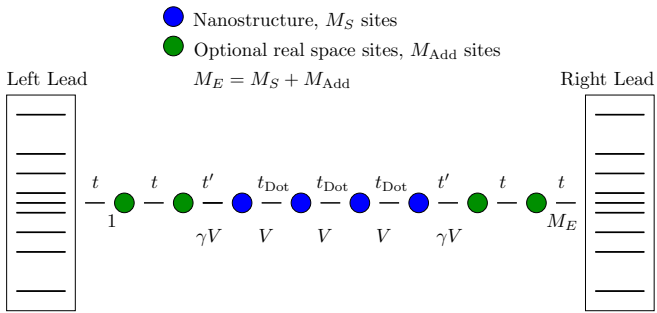}
 \caption{Schematics of the leads coupled to the nanostructure.} 
 \label{fig:SL}       
\end{figure}

In order to be able to describe processes on different energy scales we first couple
our nanostructure to a few sites in real space to capture local, i.e.\ high energy, physics.
Then we employ a logarithmic discretization of the momentum leads to cover a large energy range
and finally we use a linear discretization of the low energy scale in order to describe low energy
transport properties accurately. We would like to note that these additional sites on a linear discretization
close to the Fermi edge are beyond a NRG like description. While they are not needed for a qualitative
description, they enable us to get very accurate results even close to the resonant tunneling regime.
The reason for that lies in the nature of transport properties, where the $\eta$ in the correlation
function plays a much more important role than for equilibrium properties. It does not only provide a smoothing
of the poles, it has to create excitations which then can actually lead to transport.

The models considered in this work are the interacting resonant
level model (IRLM) and the natural extension of this model to linear
chains, defined by the Hamiltonians
\begin{eqnarray}
    H_{RS} &=& \sum_{j\in S} \mu_g \hat{c}_j^\dagger  \hat{c}^{}_j
       \,-\, \sum_{j,j-1\in S_E} \big(t_j  \hat{c}_j^\dagger \hat{c}^{}_{j-1}+ \text{h.c.}\big)\nonumber\\
    && +\, \sum_{j,j-1\in S_E} V_j \Big( \hat{n}_j-\frac 12\Big) \Big( \hat{n}_{j-1} - \frac 12\Big),\\
    H_{MS} &=& \sum_{k\in L,R} \epsilon^{}_k \hat{c}_k^\dagger  \hat{c}^{}_k,\\
    H_{T} &=& -t_k \Big( \sum_{k\in L}\hat{c}_{k}^\dagger  \hat{c}^{}_{1}
            \,+\,   \sum_{k\in R} \hat{c}_{k}^\dag  \hat{c}^{}_{M_E} \Big) \,+\, \text{h.c.},
\end{eqnarray}
%%    H_{SL} &=& -t(\hat{c}_{L_1}^\dagger  \hat{c}_{1} + h.c.) -t( \hat{c}_{R_1}^\dag
%%     \hat{c}_{M_E} + h.c.) \,,
where $\hat{c}^\dag_\ell$ and $\hat{c}^{}_\ell$ 
($\hat{c}^\dag_k$ and $\hat{c}^{}_k$) are the spinless
fermionic creation and annihilation operators at site $\ell$
(momentum $k$),
$\hat{n}_\ell = \hat{c}^\dag_\ell \hat{c}^{}_\ell$. $H_{RS}$,
$H_{MS}$, and $H_T$ denote real space, momentum space, and tunneling
between real- and momentum space Hamiltonians respectively. The
symbols $S$ and $S_E$ denote the nanostructure and the extended
nanostructure (the full real space chain) respectively.  The
indices $1$ and $M_E$ denote the first and last site in $S_E$. The
general setup and the specific values of the hopping matrix elements
$t_j$ and the interactions $V_j$ are indicated in
Fig.~\ref{fig:SL}, and note specifically the interactions on
the contact links, $\gamma V$. The momentum dependent coupling $t_k$
is chosen to represent an infinite onedimensional tight-binding chain
if a cosine band $\epsilon_k  = -2 t \cos(k)$ is chosen. All energies are measured in
units of $t=1$.

\begin{figure}
 \centering
 \includegraphics[width=\columnwidth]{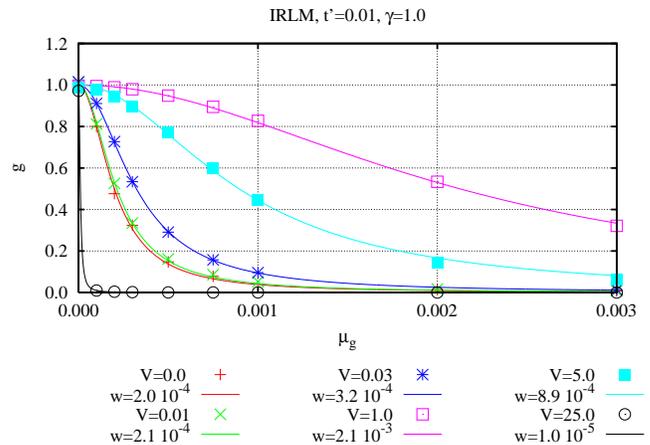}
 \caption{Linear conductance versus gate potential for the interacting resonant level model for $t'=0.01$
    and a interaction on the contacts ranging from zero to 25. 
    To each set of DMRG data a Lorentzian of half width $2$w has been added as a guide to the eye.
    The leads are described with a cosine band between $\pm2$ such that the Fermi velocity is $v_F=2$.
    In contrast to intradot interaction the   interaction on the contacts  enhances the conductance 
    and shows a non monotonic behavior versus contact interaction.
} 
 \label{fig:G1}       
\end{figure}

In fig.~\ref{fig:G1} we show the linear conductance versus gate potential for a contact hopping of
$t'=0.01$ and interaction on the contacts ranging from zero to 25.
The calculations have been performed with 130 sites in total,
$M_E=10$ real space sites, and 120 momentum space sites. Due
to the symmetry of the band we used a discretization that is
symmetric around $\epsilon_F=0$, and applied an identical
discretization scheme to both leads. To represent the `large' energy
span in the band we used 20 logarithmically scaled sites, and
thereafter used 10 linearly spaced sites to represent the low energy
scale correctly. In the DMRG calculations presented we used at least
1300 states per block and 10 finite lattice sweeps.

The data demonstrates a strong increase of the resonance width due to interaction up to
a factor of ten. The increase of the resonance width due to  interaction on the contact is in
contrast to the reduction of conductance due to interaction on nanostructures, see \cite{BSW06}.
Once interaction is larger than the Fermi velocity the resonance width gets strongly reduced. 
The results also shows that we can now resolve resonance width of the order $10^{-5}$. 
We would like to note that this scheme is not restricted to single impurity models and that 
it also works for extended nanostructures.

The implementation of this new scheme was only finished recently and we are currently extending it to 
include the spin degree of freedom. In detail we study the single impurity Anderson model
attached to polarized, ferromagnetic leads.

\section{Further Projects}
The code developed within this project has also been used in \cite{Molina_Schmitteckert:2007} to
study quantum phase transition with entanglement entropy
and in \cite{Molina_Dukelsky_Schmitteckert:2007} to study onedimensional
fermions in a harmonic trap with an attractive on site interaction.
In \cite{Schmitteckert_Evers:2007} we used the DMRG to extract the exact corresponding
functionals of a lattice Density Functional Theory and compared the conductance calculations 
within DMRG and DFT.

\section{Post-Publication Note}
In the code applied within this report, some off-diagonal blocks where missing in the 
construction of the reduced density matrix for the time-dependent simulation.
In return the selected basis is sub-optimal. We checked with a corrected code version,
that the induced error is indeed small, especially for the 5000 states.
The content of the report is not affected. The results of our td-DMRG variant could 
just have been a little bit better.
%%
%%
%%%%%%%%%%%%%%%%%%%%%%%%%%%%%%%%%%%%%%%%%%%%%%%%%%%%%%%%%
\section*{Acknowledgments}
%%%%%%%%%%%%%%%%%%%%%%%%%%%%%%%%%%%%%%%%%%%%%%%%%%%%%%%%%
%%
%%
We would like to thank Corinna Kollath for interesting discussions and for
providing us with the raw data of \cite{Kollath_Schollwoeck_Zwerger:2005}.
This work profited from the parallelization performed within the project 710
of the Landesstiftung Baden-Württemberg. The reformulation of the momentum leads
was performed together with Dan Bohr within the HPC-EUROPA project RII3-CT-2003-506079.
Most of the calculations have been performed at the XC1
and XC2 of the SSC Karlsruhe under the grant number RT-DMRG.

\end{document}